\documentclass[twocolumn,superscriptaddress,showpacs,aps,floatfix]{revtex4-1}
\usepackage{amsmath,amsfonts,amssymb,epsfig,dcolumn,bm,dsfont,graphics,latexsym,color,graphicx,multirow,ulem}

\let\oldAA\AA

\renewcommand{\AA}{\text{\normalfont\oldAA}}
\newcommand{\ket}[1]{| {#1} \rangle} 
\begin{document}
\preprint{AIP/123-QED}
\title{Proximity-enabled control of spin-orbit coupling in phosphorene symmetrically and asymmetrically encapsulated by WSe$_2$ monolayers}
\author{Marko Milivojevi\'c}
\affiliation{Institute of Informatics, Slovak Academy of Sciences, 84507 Bratislava, Slovakia}
\affiliation {Faculty of Physics, University of Belgrade, 11001 Belgrade, Serbia}
\author{Martin Gmitra}
\affiliation{Institute of Physics, Pavol Jozef \v{S}af\'{a}rik University in Ko\v{s}ice, 04001 Ko\v{s}ice, Slovakia}
\affiliation{Institute of Experimental Physics, Slovak Academy of Sciences, 04001 Ko\v{s}ice, Slovakia}
\author{Marcin Kurpas}
\affiliation{Institute of Physics, University of Silesia in Katowice, 41‑500 Chorz\'ow, Poland}
\author{Ivan \v Stich}
\affiliation{Institute of Informatics, Slovak Academy of Sciences, 84507 Bratislava, Slovakia}
\affiliation{Department of Natural Sciences, University of Saints Cyril and Methodius, 917 01 Trnava, Slovakia}
\author{Jaroslav Fabian}
\affiliation{Institute for Theoretical Physics, University of Regensburg, 93053 Regensburg, Germany}
\begin{abstract}
We analyze, using first-principles calculations and the method of invariants, the spin-orbit proximity effects in trilayer heterostructures comprising phosphorene and encapsulating WSe$_2$ monolayers. We focus on four different configurations, in which the top/bottom WSe$_2$ monolayer is twisted by 0 or 60 degrees with respect to phosphorene, and analyze the spin splitting of phosphorene hole bands around the $\Gamma$ point. 
Our results show that the spin texture of phosphorene hole bands can be dramatically modified by different encapsulations of phosphorene monolayer.
For a symmetrically encapsulated phosphorene, the momentum-dependent spin-orbit field has the out-of-plane component only, simulating the spin texture of phosphorene-like group-IV monochalcogenide ferroelectrics. Furthermore, we reveal that the direction of the out-of-plane spin-orbit field can be controlled by switching the twist angle from 0 to 60 degrees. Finally, we show that the spin texture in asymmetrically encapsulated phosphorene has the dominant in-plane component of the spin-orbit field, comparable to the Rashba effect in phosphorene with an applied sizable external electric field. Our results confirm that the significant modification and control of the spin texture is possible in low common-symmetry heterostructures, paving the way for using different substrates to modify spin properties in materials important for spintronics. 
\end{abstract}
\maketitle
\section{Introduction}
The electron spin can add new functionalities to electronic devices~\cite{W+01,ZFS04,FME+04}. 
Although graphene is a promising material for spintronics applications~\cite{HKG+14} due to its extraordinary electron mobility~\cite{TJP+07}, the lack of a bandgap
restricts its exploitation in semiconducting spin devices.
Besides graphene, different semiconducting two-dimensional materials~\cite{FKR+99,OCM+00,J12}, including phosphorene~\cite{ATK+17} have been considered a potential platform for spintronics.

Electric spin control and manipulation are facilitated by spin-orbit coupling (SOC)~\cite{NTA+20,M21}. To induce or enhance SOC in a desired material, different approaches have been used, such as the electric-field induced Rashba SOC~\cite{GKE+09} or transfer of SOC between materials in van der Waals (vdW) heterostructures via the proximity effect~\cite{SFK+21}.

In materials with low atomic $Z$ number, therefore with weak intrinsic SOC, electrical tuning of SOC is not very efficient~\cite{SPS14,KGF16}. Instead, the proximity effect emerges as the most viable direction for inducing sizable SOC in low Z materials. Using different vdW heterostructures of graphene and strong SOC/magnetic materials, significant progress towards manipulation of graphene's spin~\cite{GF15,GF17,ZGF19,ZCR+21,SMK+23}, and magnetic~\cite{ZGF+16,ZGF18,ZJF19,ZGF20,ZGF22,ZF22} properties has been demonstrated.
Focusing on phosphorene~\cite{LYY+14,RCN14,LNZ+14,FGF+15,LDD+15,LKJ+17}, a material whose sizable semiconducting gap~\cite{SMG+16,ZHC+17,FDT+19,HFM+23} makes it a promising spintronics material, a similar approach can be 
used. The obvious first choice is the vdW heterostructure made of phosphorene and a transition-metal dichalcogenide (TMDC)~\cite{MGK+23} monolayer (ML), due to the huge spin splitting in the valence bands of TMDC materials~\cite{ZCS11,KZD+13,SYZ+13,ABX+14,KBG+15}. 
The low symmetry of the phosphorene/TMDC heterostructure imposes minimal restrictions on the phosphorene's spin texture, suggesting there is an immense possibility for manipulation of the SOC in phosphorene through twisting and encapsulation.
\begin{figure*}[t]
\centering
\includegraphics[width=0.99\textwidth]{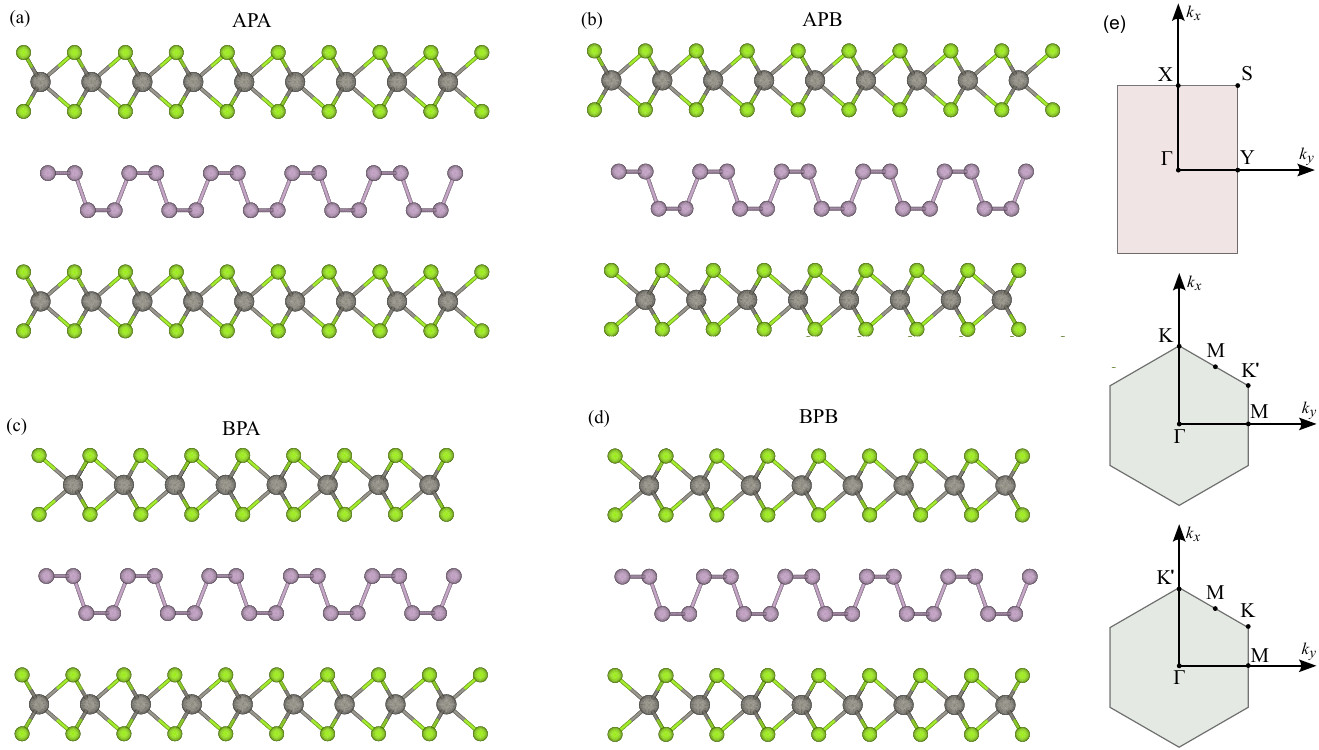}
\caption{Side view of the atomic structure models (a)-(d) of four different WSe$_2$/P/WSe$_2$ heterostructures. We label different heterostructures as IPJ, $I,J=A,B$, where I/J corresponds to the top/bottom WSe$_2$ ML, while the label
A/B corresponds to the relative twist angle of $0/60$ degrees of the WSe$_2$ ML with respect to the phosphorene ML.  Furthermore, the $x$/$y$ direction of the heterostructure corresponds to the zigzag/armchair direction of the phosphorene ML. In~(e) the Brillouin zone with high symmetry points of phosphorene and A and B WSe$_2$ MLs is given.}\label{FigHet}
\end{figure*} 

In Ref.~\cite{MGK+23}, it was shown that the reduced symmetry of phosphorene from ${\bm D}_{2h}$~\cite{BC72,LA14} to ${\bm C}_{1{\rm v}}$ within the
$0^{\rm o}/60^{\rm o}$ twist angle phosphorene/WSe$_2$ heterostructure triggers two types of spin-orbit fields in phosphorene holes close to the $\Gamma$ point: the in-plane which is a consequence of the broken nonsymmorphic horizontal mirror plane symmetry and the out-of-plane, triggered by the broken rotational symmetry. Based on the first principles calculations and symmetry analysis, we show that phosphorene encapsulated by two WSe$_2$ MLs
can be used to additionally control and manipulate the spin texture and SOC strength in phosphorene holes. More specifically, we demonstrate the increase in the out-of-plane spin-orbit field by symmetrical encapsulation of the phosphorene ML, whereas in phosphorene asymmetrically encapsulated by two WSe$_2$ MLs a sizable strength of the in-plane spin-orbit field can be achieved, comparable with the phosphorene modulated by a strong external electric field~\cite{KGF16}. 
In both cases, the encapsulation by WSe$_2$ modifies the anisotropy of spin splitting in the phosphorene monolayer. For symmetrically encapsulated phosphorene, we observe a giant anisotropy of splitting along the main crystallographic direction, in contrast to the symmetrical encapsulation, for which spin splitting is only slightly anisotropic.

This paper is organized in the following way. After a short introduction, in Sec.~\ref{Heterostructure} we analyze the geometry of the heterostructure in which phosphorene is symmetrically/asymmetrically encapsulated by WSe$_2$ MLs and provide numerical details relevant to the first-principles calculation of the band structure. Based on the band structure calculations in the following Sec.~\ref{Bandstructure}, we analyze the spin-orbit coupling of phosphorene holes close to the $\Gamma$ point and extract the relevant SOC parameters for each heterostructure studied. 
Finally, in Sec.~\ref{conclusions}, the most important conclusions are given.

\section{Atomic structure and ab-initio calculation details}\label{Heterostructure}
In Fig.~\ref{FigHet}~(a)-(d), we present side views of the atomic structure models of the WSe$_2$/P/WSe$_2$ heterostructures, with different relative orientations between the top and bottom  WSe$_2$ MLs and phosphorene. The commensurate heterostructures were constructed using the CellMatch code~\cite{L15}, containing 20 P, 16 W, and 32 Se atoms. We kept the phosphorene unstrained, at the same time straining both WSe$_2$ monolayers by 0.51\%. We label the A (B) orientation of WSe$_2$ ML having the twist angle $0^{\rm o}$ ($60^{\rm o}$) with the armchair direction of phosphorene ML. Furthermore, in Fig.~\ref{FigHet}~(e), the relative orientation of phosphorene's Brillouin zone (BZ) with respect to the BZ orientation of A and B WSe$_2$ MLs is given.
We will label the four heterostructures as IPJ, I,J=A,B,  where I/J corresponds to the top/bottom WSe$_2$ ML. We emphasize that the vertical mirror plane symmetry, present in both PA and PB bilayer heterostructures~\cite{MGK+23}, is preserved in the trilayer case since putting one more A or B-oriented WSe$_2$ monolayer on top of the bilayer heterostructure is compatible with the bilayer's symmetry.
In all four cases, the vertical mirror plane symmetry coincides with the $yz$ plane, where the 
armchair (zigzag) direction corresponds to the $y$ $(x)$ direction of the heterostructure.

Lattice vectors of phosphorene ML are equal to ${\bm a}=a {\bf e}_x$, ${\bf b}=b {\bf e}_y$,
where $a=3.2986\AA$ and $b=4.6201\AA$~\cite{JKG+19}, while the lattice vectors of WSe$_2$ ML correspond to
 ${\bm a}_1=a_{\rm W} {\bf e}_x$,  ${\bm a}_2=a_{\rm W}(-{\bf e}_x+\sqrt{3}{\bf e}_y$)/2 ($a_{{\rm W}}=3.286\AA$~\cite{WJ69}).
\begin{figure*}[t]
\centering
\includegraphics[width=0.99\textwidth]{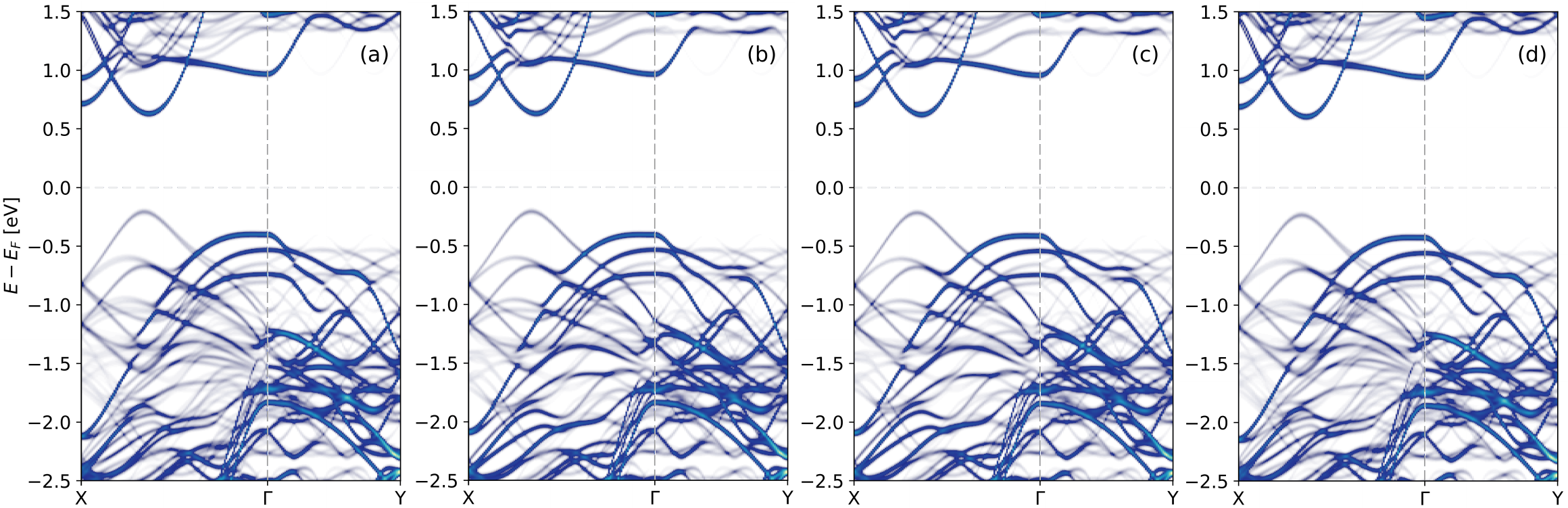}
\caption{In panels (a)-(d), the calculated band structure of APA, APB, BPA, and BPB heterostructures, respectively, unfolded to the X$\Gamma$Y path of the phosphorene Brillouin zone is presented.
}\label{FigUnfold}
\end{figure*}
DFT electronic structure calculations of the WSe$_2$/P/WSe$_2$ heterostructure were performed using the plane wave Q{\sc{uantum}} ESPRESSO (QE) package~\cite{QE1,QE2}. The Perdew–Burke–Ernzerhof exchange-correlation functional was employed~\cite{PBE96}, combined with the norm-conserving method~\cite{HSC79}. Atomic relaxation was performed using the quasi-Newton scheme and scalar-relativistic SG15 Optimized Norm-Conserving Vanderbilt (ONCV) pseudopotentials~\cite{H13,SG15,SGH+16}. For ionic minimization, the force and energy convergence thresholds were set to $1\times10^{-4}$~Ry/bohr and $10^{-7}$ Ry, respectively. Additionally, the Monkhorst-Pack scheme with $56\times 8$ $k$-points mesh was used, small Methfessel-Paxon energy level smearing of 1~mRy~\cite{MP89}, and kinetic energy cut-offs for the wave function and charge density 80\,Ry and 320\,Ry, respectively. Since the distance between the monolayers in the van der Waals heterostructures plays a key role in the spin-orbit proximity effect, we have incorporated three different van der Waals corrections: semiempirical Grimme's DFT-D2 (D2)~\cite{G06,BCF+09}, 
Tkachenko-Scheffler (TS)~\cite{TS09}, and the non-local rvv10~\cite{VV10,SGG13}. 
Finally, a vacuum of 20\,${\AA}$ in the $z$-direction was used in all cases. For the relaxed structures, the average distance between the bottom/top phosphorene and the closest selenium plane in the $z$-direction is given in 
Table~\ref{distance}. All four configurations have approximately the same energy. However, the asymmetric heterostructures are slightly more stable than the symmetric ones in terms of total energy by about $0.8-2$ meV/atom, depending on the vdW correction used, see Table~\ref{distance}.
In the case of noncolinear DFT calculations with spin-orbit coupling, fully relativistic SG15 ONCV pseudopotentials were utilized. Also, the dipole correction~\cite{B99} was applied to properly account for the energy offset due to dipole electric field effects. The energy convergence threshold for the noncolinear calculations
was set to $10^{-8}$~Ry, while the $k$-points mesh and kinetic energy cutoffs for the wave function and charge density were the same as in the relaxation calculations. 

\begin{table}[htp]
\caption{Average distance $d_{\rm PSe}^{\rm b/t}$ between the bottom/top phosphorene and the closest selenium plane in the $z$-direction for the relaxed APA, APB, BPA, and BPB heterostructures 
obtained with different vdW corrections: Grimme-D2 (listed as D2), non-local rvv10, and Tkachenko-Scheffler. Total energy difference $\Delta E$ with respect to the minimal one (set to be zero) for each vdW correction is also given.}
\label{distance}
\centering
\begin{tabular}{c|c|c|c|c}
\hline
het&vdW corr&$d_{\rm PSe}^{\rm b}$ [\AA]&$d_{\rm PSe}^{\rm t}$ [\AA] & $\Delta E$ [meV/atom]\\\hline
\multirow{3.1}{*}{APA}& D2& 3.34 &  3.35 & 1.993 \\
& rvv10& 3.45 &  3.45 & 1.090\\
& TS& 3.69 &  3.69 & 0.794 \\\hline
\multirow{3.1}{*}{APB}& D2&  3.31 &  3.31 & 0.034 \\
& rvv10& 3.42 &  3.42 & 0.022\\
& TS& 3.66 &  3.66 & 0 \\\hline
\multirow{3.1}{*}{BPA}& D2& 3.31 &  3.31 & 0 \\\
& rvv10& 3.42 &  3.42 & 0 \\
& TS& 3.66 &  3.66 & 0.006 \\\hline
\multirow{3.1}{*}{BPB}& D2& 3.35 &  3.34 & 1.992 \\
& rvv10& 3.45 &  3.45 & 1.091\\
& TS& 3.69 &  3.69 & 0.795\\\hline
\end{tabular}
\end{table}

Finally, it is to be noted that the illustration of the band structure, unfolded to the Brillouin zone of both MLs, is done using the DFT Vienna ab-initio simulation package VASP~6.2~\cite{KF96,KF99}, using as the input the relaxed structure with D2 vdW correction from QE code.

\section{Band structure analysis}\label{Bandstructure}
To understand the interaction between the phosphorene and WSe$_2$ MLs, in  Fig.~\ref{FigUnfold}~(a)-(d), we present the band structure of APA, APB, BPA, and BPB heterostructures, unfolded to the X$\Gamma$Y path of the phosphorene monolayer Brillouin zone.
As can be seen,
the band structures are almost identical. This is not surprising, since twisting the WSe$_2$ ML by 60 degrees interchanges the K and K' points [see Fig. \ref{FigHet} (e)]. This further means that the $\Gamma$K/$\Gamma$K' line of the WSe$_2$  BZ is projected onto the $\Gamma$X line of the phosphorene BZ for the 0/60 degrees twist angle; in the case of the 
$\Gamma$Y line of the phosphorene BZ, the $\Gamma$M line of the WSe$_2$ BZ is projected.
However, although the energies of WSe$_2$ at the $\Gamma$K and $\Gamma$K' lines are identical, the corresponding spin expectation values are swapped. This can be illustrated in the example of the equal energies at the K and K' points, connected by the time-reversal symmetry $\Theta$ through the relation $\Theta E_{\ket{{K}+}}=E_{\ket{{K'}-}}$, where $\ket{\pm}$ is the spin wavefunction with $s_z=\pm 1/2$ spin expectation value. 

To obtain the quantitative estimate of the spin-orbit proximity effect in phosphorene, we will focus on the spin-splitting of the top valence band around the $\Gamma$ point that has the dominant phosphorene character. The phosphorene character can be simply identified as a strong asymmetry of the band dispersion in different directions around the $k=0$ point~\cite{FY14}. Furthermore, the previous analysis shows that two qualitative distinct situations occur: in the case of symmetric encapsulation, APA (BPB), the K (K') points of the top and bottom WSe$_2$ ML are positioned on the $\Gamma X$ line, while for the asymmetric encapsulation, APB (BPA), the K point of the top (bottom) WSe$_2$ ML is followed by the K' point of the bottom (top) WSe$_2$ ML.

\subsection{Effective parameters}\label{effectivepar}
The influence of different encapsulations on the spin-orbit proximity effect of phosphorene holes can be quantitatively described in terms of the effective spin-orbit Hamiltonian model,
consistent with the ${\bf C}_{1{\rm v}}$ symmetry of the studied heterostructures. As shown in~\cite{MGK+23},
the spin-orbit splitting of phosphorene holes in the vicinity of the $\Gamma$ point can be expressed in terms of the linear-in-momenta spin-orbit field
\begin{equation}\label{effectiveHam}
 H_{\rm eff}=\lambda_1 k_x \sigma_y+\lambda_2 k_y \sigma_x + \lambda_3 k_x \sigma_z,
\end{equation}
where $\sigma_{i}$, $i=x,y,z$, represents components of the Pauli matrix operator ${\bm \sigma}$ which is connected to the spin operator ${\bm S}$ via the relation ${\bm S}=(\hbar/2){\bm \sigma}$, while the parameters $\lambda_1$, $\lambda_2$, and $\lambda_3$ need to be determined for each stacking.
In terms of the spin-orbit field that can be induced via the proximity effect, one can consider the $\lambda_1 k_x \sigma_y+\lambda_2 k_y \sigma_x$ terms as the in-plane spin-orbit field, triggered by breaking the horizontal glide mirror plane symmetry of the phosphorene ML. 
On the other hand, the $\lambda_3 k_x \sigma_z$ term represents the out-of-plane spin-orbit field, present due to the broken twofold out-of-plane rotational symmetry.
\begin{figure*}[t]
\centering
\includegraphics[width=0.75\textwidth]{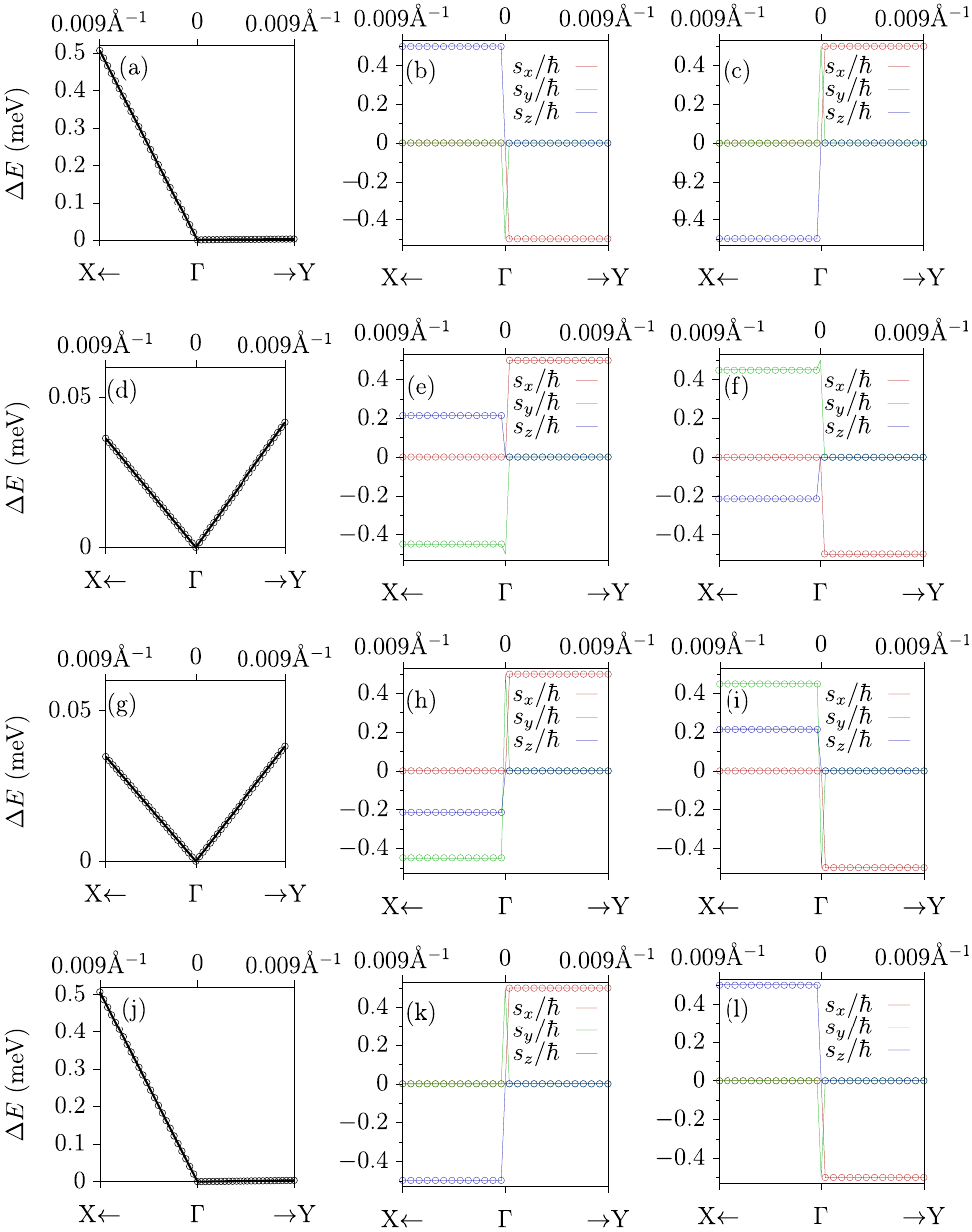}
\caption{Comparison between the DFT data and the spin-orbit Hamiltonian model with SOC parameters from Table~\ref{vdWinfluence}. More detailed, in (a)-(c), (d)-(f), (g)-(i), and (j)-(l), spin-orbit induced band splitting $\Delta E$ between the spin-split upper and lower top valence band around the $\Gamma$ point, as well as the corresponding spin expectation values for the heterostructures APA, APB, BPA, and BPB, correspondingly. In all four cases, the Grimme-D2 vdW correction is used for the relaxation of the heterostructure,  while the $k$-path follows the high symmetry lines X$\Gamma$ ($k_x\in(\kappa,0)$) and $\Gamma$Y ($k_y\in(0,\kappa)$), 
where $\kappa=0.009~{\rm\AA}^{-1}$.}\label{comparisonFIT}
\end{figure*}

\begin{figure}
    \centering    \includegraphics[width=\columnwidth]{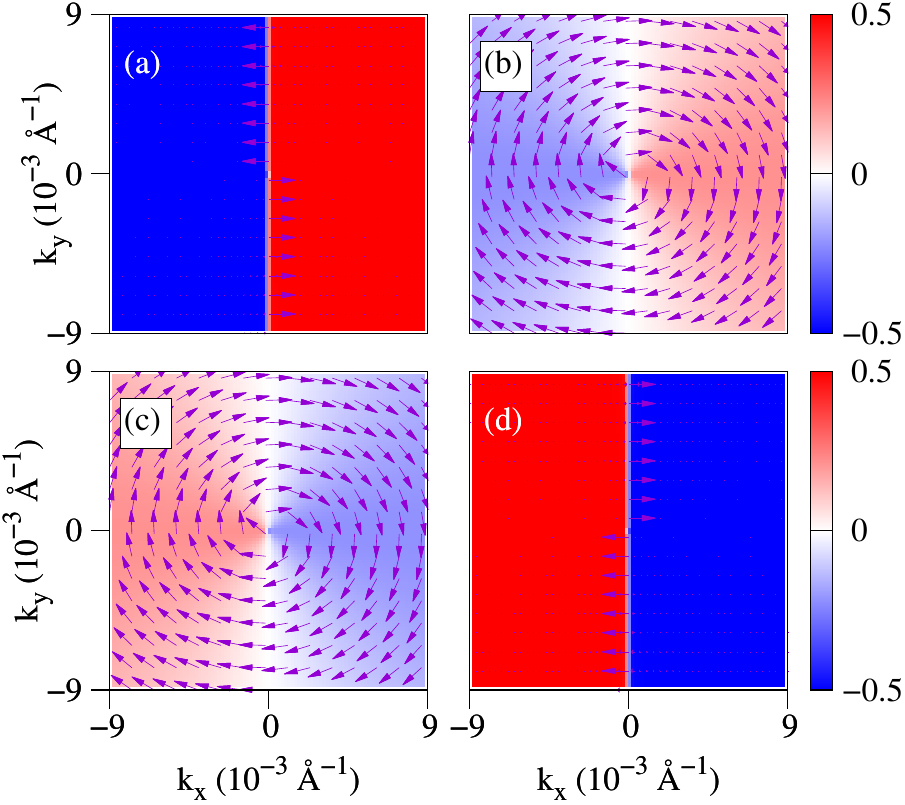}
    \caption{
    Spin textures of the lower spin split top valence band branch around the $\Gamma$ point ($k_x/k_y\in(-\kappa,\kappa)$, $\kappa=0.009~{\rm\AA}^{-1}$) for APA~(a), APB~(b), BPA~(c), and BPB~(d) heterostructures, obtained using the data from Table~\ref{vdWinfluence} (D2 case). The in-plane spin texture is represented by arrows, while the $z$ component of the spin is given by color scale.}
    \label{spinTEXTURE}
\end{figure}

The obtained spin-orbit parameters are gathered in Table~\ref{vdWinfluence}, obtained by fitting the spin-orbit Hamiltonian model to DFT data. In more detail, the spin-splitting of the top valence band of phosphorene around the $\Gamma$ point and spin-expectation values are fitted to the model, assuming the $X\Gamma$, $k_x\in(-\kappa,0)$, and $\Gamma Y$ path, $k_y\in(0,\kappa)$, where $\kappa=0.009\,\AA^{-1}$.
The comparison between the DFT data and the model in the case of the Grimme-D2 vdW correction is given in Fig.~\ref{comparisonFIT}, for considered heterostructures, while the corresponding spin texture of the lower spin split branch is given in Fig.~\ref{spinTEXTURE}. Concerning the lower branch, the spin texture of the upper branch has the opposite sign, ${\bf s}({\bf k})_{\rm upper}=-{\bf s}({\bf k})_{\rm lower}$, where ${\bf s}({\bf k})$ represents the spin expectation values of the operator ${\bf S}$, ${\bf s}({\bf k})=\langle {\bf S}({\bf k}) \rangle$, at a ${\bf k}$ point of the Brillouin zone.

What can be concluded from the effective spin-orbit parameters? First, we notice that in symmetric IPI heterostructures, the in-plane spin-orbit field is strongly suppressed, while the intensity of the out-of-plane spin-orbit field is enhanced roughly twice with respect to the P/WSe$_2$ heterostructure~\cite{MGK+23}.
Also, giant anisotropy of the spin splitting between the $\Gamma$X and $\Gamma$Y directions can be noticed, being proportional to $\sqrt{\lambda_1^2+\lambda_3^2}/|\lambda_2|$. 
On the other hand, in the asymmetric IPJ heterostructures, the out-of-plane spin-orbit field is strongly suppressed, whereas the in-plane spin-orbit field is much stronger than in symmetric IPI heterostructures, although still a few times weaker than in the P/WSe$_2$ bilayer~\cite{MGK+23}. The spin-splitting anisotropy is much less pronounced; by calculating $\sqrt{\lambda_1^2+\lambda_3^2}/|\lambda_2|$ for Grimme-D2, rvv10, and TS vdW corrections we get 0.87, 1.52, and 3.15 in the APB case and 0.91, 1.63, and 3.77 in the BPA case, respectively. Although the size of the spin splitting in the asymmetric case is similar to the Rashba effect of pristine phosphorene~\cite{KGF16}, the ratio of the spin splitting between the $\Gamma$X and $\Gamma$Y directions in pristine phosphorene in an electric field is equal to 1/3~\cite{KGF16}, showing different trends than in asymmetrically encapsulated phosphorene. 

The above discussion leads to the simple conclusion that an in-plane spin-orbit field prefers asymmetric encapsulation, whereas an out-of-plane spin-orbit field can be boosted by symmetric encapsulation. 
\begin{table}[t]
\caption{Extracted SOC parameters $\lambda_{\rm 1/2/3}$ of the top valence band of phosphorene around the $\Gamma$ point for APA, APB, BPA, and BPB heterostructures, obtained using different vdW corrections: Grimme-D2 (listed as D2),  non-local rvv10, and Tkachenko-Scheffler.}\label{vdWinfluence}
\centering
\small
\setlength{\tabcolsep}{7pt}
\renewcommand{\arraystretch}{1.0}
\begin{tabular}{c|c|c|c|c}
\hline
het&vdW & $\lambda_{1}$ [meV\,\AA]&$\lambda_{2}$ [meV\,\AA]&$\lambda_{3}$ [meV\,\AA]\\\hline
\multirow{3.1}{*}{APA}&D2 &0.00 &  0.12 & -27.68\\
&rvv10 &0.00 &  0.03 & -20.27\\
&TS &0.01 &  0.04 & -9.48\\\hline
\multirow{3.1}{*}{APB}&D2    &1.79 &  -2.27 & -0.86\\
&rvv10 &0.76 &  -0.59 & -0.48\\
&TS    &0.14 &  -0.07 & -0.17\\\hline
\multirow{3.1}{*}{BPA}&D2    &1.72 &  -2.09 &  0.82\\
&rvv10 &0.79 &  -0.56 &  0.46\\
&TS    &0.16 &   0.06 &  0.16\\\hline
\multirow{3.1}{*}{BPB}&D2    &0.00 &  -0.19 & 27.72\\
&rvv10 &0.00 &  -0.27 & 20.31\\
&TS &0.00 &  0.01 & 9.47\\\hline
\end{tabular}
\end{table}
This conclusion is consistent with the simple argument that the broken horizontal mirror plane symmetry triggers the in-plane spin-orbit field and is a measure of the structural inversion asymmetry in the $z$-plane which is the highest in the bilayer heterostructures, still present in asymmetric encapsulation, and nearly absent in symmetric encapsulation of phosphorene. On the other hand, the effective out-of-plane spin-orbit field
originates from the broken out-of-plane rotational symmetry and can be interpreted as an effective in-plane electric field in the $y$-direction, $ {\bf E}=E {\bf e}_y$, giving rise to the Rashba SOC term $(E {\bf e}_y\times   {\bf k})\cdot {\bm \sigma}\propto k_x \sigma_z$. In the case of P/WSe$_2$ heterostructures, it was shown that the twist angle of WSe$_2$ can control the sign of this term, changing the value of $\lambda_3$ from negative to positive with a twist angle switch from 0$^{\rm o}$ to 60$^{\rm o}$~\cite{MGK+23}.  

One can expect, that in a trilayer heterostructure, the sign and value of $\lambda_3$ will depend on the relative orientation of the top and bottom WSe$_2$ layers.  
In the case of the symmetric APA (BPB) heterostructure, the overall $\lambda_3$ should be negative (positive) with the intensity roughly doubled in comparison to the P/A (P/B) bilayer heterostructure~\cite{MGK+23}, since both layers contribute with the same sign (at the same K/K' valleys). 
In the case of the asymmetric IPJ heterostructure, the contributions to $\lambda_3$ from the top (K) and bottom (K') WSe$_2$ layers should cancel out due to opposite spin polarization at the K and K' valleys, giving in effect a small overall value of $\lambda_3$. 
This conjecture is confirmed by the spin texture shown in Fig.~\ref{spinTEXTURE}. For symmetrically encapsulated phosphorene, the  $s_z$ spin component is close to 0.5 (in $\hbar$ unit) and takes opposite values for APA and BPB heterostructures, see Fig.~\ref{spinTEXTURE}(a) and (d), respectively. The in-plane spin components are almost zero, except $s_y$ for $k_x\approx 0$. For APB and BPA heterostructures, Fig. \ref{spinTEXTURE}(c), (d),  $s_z$ is close to zero, which is reflected in the small values of $\lambda_3$ collected in Table~\ref{vdWinfluence}.

Finally, we notice that the choice of the vdW correction has a quantitative influence on the obtained results, but the overall qualitative picture remains intact (see Table~\ref{vdWinfluence}). The relaxation procedure has revealed that within the three chosen vdW corrections the Grimme-D2 gives the minimal distance between the bottom/top phosphorene plane and the closest selenium plane in the $z$-direction, while in the case of TS vdW correction, the distances are the biggest.
This is straightforwardly reflected on $\lambda$ parameters calculated in each case, since the distance between the phosphorene and selenium plane has an exponential impact on the virtual transitions between the phosphorene and bottom/top WSe$_2$ ML, being the microscopic mechanism of the SOC transfer between the MLs~\cite{DRK+19}. Also, it is to be noted that the SOC parameters were unaffected when the self-consistent calculation was performed with and without the vdW correction, suggesting the crucial role of vdW correction in the relaxation procedure and the need for vdW correction benchmarking of ML distances in various heterostructures important for spintronics application and beyond.

\subsection{Proximity-induced spin relaxation}\label{sec:relaxation}
Let us now discuss the possible implications of the proximity-induced SO fields for spin relaxation in phosphorene.  Pristine phosphorene is characterized by an extraordinarily long, nanosecond-range spin lifetime \cite{ATK+17,CLT+24}. The upper limit for the lifetime comes from the Elliott-Yafet (EY) mechanism \cite{EY1954}, while the Dyakonov-Perel (DP) mechanism \cite{DP1971} is less important, provided that the extrinsic Rashba SO, induced by the presence of a substrate or an external transverse electric field, is weak ~\cite{KGF16}. 
As we have shown above, encapsulation breaks almost all symmetries of phosphorene and activates Rashba SO fields, different for different twist angles. It is thus natural to expect differences in spin relaxation for symmetrically and asymmetrically encapsulated phosphorene. 

\begin{figure}
    \centering
    \includegraphics[width=\columnwidth]{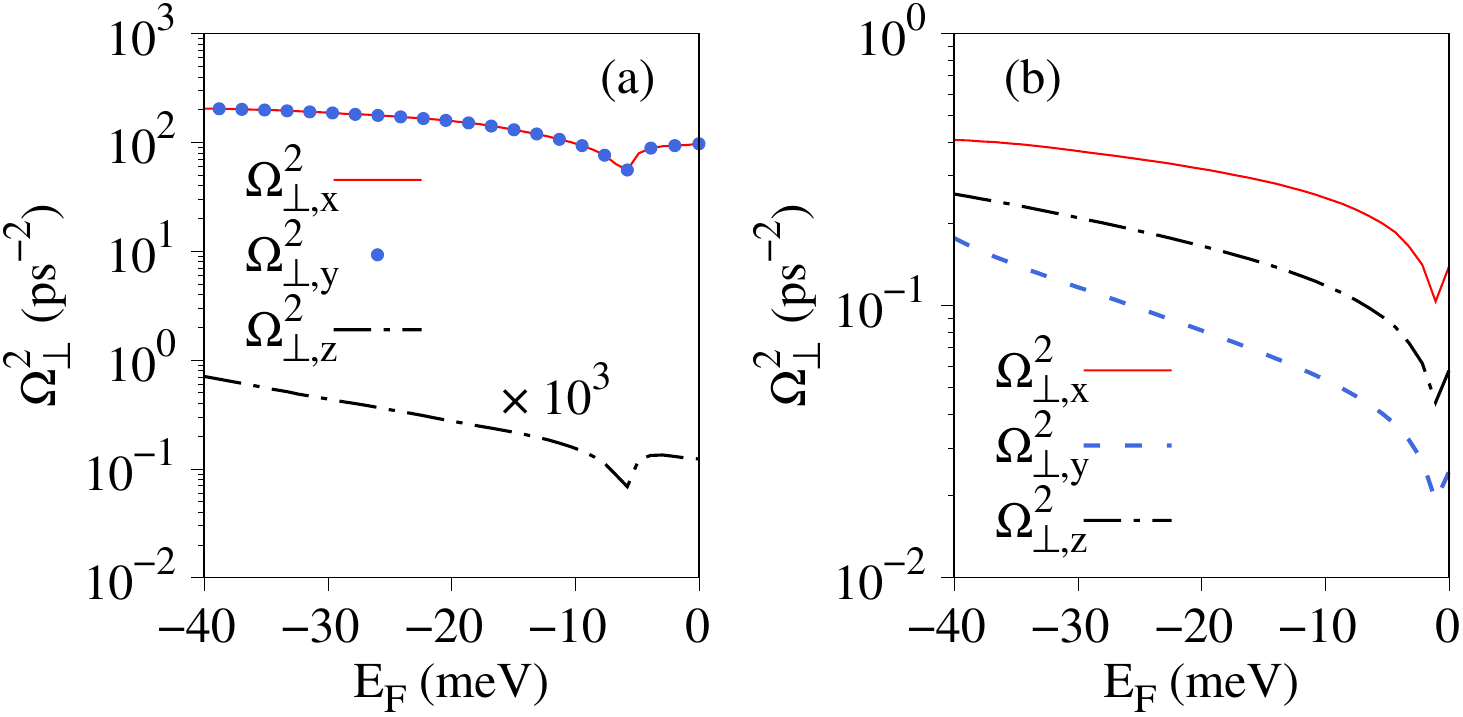}
    \caption{Fermi contour average squared spin-orbit field $\Omega^2_{\perp,i}$ perpendicular to the $i$-th spin component, $i=\lbrace x,y,z\rbrace$ versus the position of the Fermi level measured from the top of the valence band;  (a) APA stacking,  (b) APB stacking. The heterostructures were relaxed using the Grimme-D2 vdW correction. }
    \label{fig:DP_time}
\end{figure}

In the case of symmetric encapsulation, the lack of the in-plane SO field should lead to long spin coherence for $s_z$-polarized spins, which usually are unaffected by the out-of-plane SO field.  In the Dyakonov-Perel mechanism, the $s_z$ component dephases via the interaction with the perpendicular components of the extrinsic spin-orbit field $\mathbf{\Omega}$, that is via $\Omega_{x}$ and $\Omega_y$. The spin dephasing rate can be estimated as $\tau_{s,z}^{-1} \approx \tau_p \Omega_{\perp,z}^2$, where $\Omega_{\perp,z}^2 = \langle\Omega^2\rangle-\langle\Omega_z^2\rangle$ and $\langle \rangle$ denote the Fermi contour average \cite{ZFS04}, and $\tau_p$ is the momentum lifetime. As can be seen from Fig. \ref{fig:DP_time} (a), $\Omega_{\perp,z}^2$ varies between $10^{-4}$~ps$^{-2}$ and $10^{-3}$~ps$^{-2}$, which translates to DP spin lifetimes between 1~ns and 10~ns, assuming $\tau_p =0.1$~ps. This indicates, that for the APA configuration the EY and DP mechanisms have comparable contributions to spin relaxation. 

The peculiar $s_z$ spin texture of phosphorene bands around the $\Gamma$ point, see Fig.~\ref{spinTEXTURE}(a) and Fig.~\ref{spinTEXTURE}(d), creates specific conditions for spin scattering, under which the spin-texture-induced spin relaxation mechanism acts similarly to the EY. 
It happens when momentum scattering involves transitions from an initial state with $\psi_{\rm i}(k_x, k_y)$ to a final state with $\psi_{\rm f}(-k'_x, k'_y)$.  Due to the opposite spin polarization of the initial and final states, momentum scattering is accompanied by a spin flip, $\psi_{\rm i}(k_x, k_y,\sigma) \rightarrow \psi_{\rm f}(-k'_x, k'_y,-\sigma)$. More intense momentum scattering leads to more effective spin relaxation and the spin relaxation rate $\tau^{-1}_s$ follows the momentum relaxation rate, $\tau^{-1}_p$, $\tau^{-1}_s\sim \tau^{-1}_p$, which is the dependence typical to the EY mechanism. Scattering between different $k_y$ momenta without a change in $k_x$ will not change the spin, reducing the total spin-flip probability by half on average, assuming isotropic momentum scattering.  The effectiveness of this mechanism, however, can be reduced by applying a small bias voltage along the $k_y$ direction, which should limit the $k_x \rightarrow -k_x$ scattering. 
The in-plane spin components should relax incomparably faster than the $s_z$. Due to the large $\Omega^2_{\perp,x/y}$, $\Omega_{\perp,x/y} \tau_p \approx 1$ and the $s_{x/y}$ spin lifetime is limited by $\tau_p$. 

For asymmetric encapsulation, the typical motional narrowing regime of the DP mechanism applies, since for all spin-orbit field components $\Omega_{\perp,i} \tau_p \ll 1$, assuming $\tau_p=0.1$~ps, see Fig.~\ref{fig:DP_time}(b).  In this case, the estimated spin lifetimes are on the order of 0.1~ns, which is approximately ten times shorter timescale than for the EY mechanism. 
\section{Conclusions}\label{conclusions}
We studied the spin-orbit proximity effect in phosphorene symmetrically and asymmetrically encapsulated by two WSe$_2$ monolayers. Focusing on the phosphorene holes around the $\Gamma$ point, we were able to 
quantitatively describe its spin physics using the spin-orbit Hamiltonian model based on the 
${\bf C}_{1{\rm v}}$ symmetry of the studied heterostructures.
We show that the overall spin-orbit field, which can be described as a sum of the in-plane and out-of-plane spin-orbit field, can be highly tunable using the studied trilayer heterostructures. Using the symmetric encapsulation, we were able to generate the spin texture consisting of the out-of-plane spin-orbit field solely, mimicking the spin texture of phosphorene-like group-IV monochalcogenide ferroelectrics. Furthermore, we have demonstrated the sign switch of the induced out-of-plane spin-orbit field by changing the twist angle between the phosphorene and WSe$_2$ monolayer from 0 to 60 degrees. Finally, we showed that in asymmetric heterostructures, the spin texture of phosphorene hole bands has the dominant in-plane component of the spin-orbit field, comparable to the Rashba effect in phosphorene with an applied sizable external electric field.  We demonstrated significant modification and control of the spin texture in phosphorene-based heterostructures, motivating the research of different low common-symmetry heterostructures as a platform for control of spin, magnetic, and/or exciton properties in materials important for spintronics. 
\\

\acknowledgments
M.M. acknowledges the financial support
provided by the Ministry of Education, Science, and Technological Development of the Republic of Serbia. This project has received funding from the European Union's Horizon 2020 Research and Innovation Programme under the Programme SASPRO 2 COFUND Marie Sklodowska-Curie grant agreement No. 945478.
M.G.~acknowledges financial support provided by Slovak Research and Development Agency provided under Contract No. APVV-SK-CZ-RD-21-0114 and by the Ministry of Education, Science, Research and Sport of the Slovak Republic provided under Grant No. VEGA 1/0105/20 and Slovak Academy of Sciences project IMPULZ IM-2021-42 and project FLAG ERA JTC 2021 2DSOTECH.
M.K.~acknowledges financial support provided  by the National Center for Research and Development (NCBR) under the V4-Japan project BGapEng V4-JAPAN/2/46/BGapEng/2022. 
I.{\v S}~acknowledges financial support by projects APVV-21-0272, VEGA 2/0070/21, VEGA 2/0131/23, and by H2020 TREX GA No. 952165 project.
J.F. acknowledges support from Deutsche Forschungsgemeinschaft (DFG, German Research Foundation) SFB 1277 (Project-ID 314695032, project B07), SPP 2244 (Project No. 443416183), and of the European Union Horizon 2020 Research and Innovation Program under Contract No. 881603 (Graphene Flagship) and FLAG-ERA project 2DSOTECH.
The authors gratefully acknowledge the Gauss Centre for Supercomputing e.V. 
for funding this project by providing computing time on the GCS Supercomputer SuperMUC-NG at Leibniz Supercomputing Centre.

\end{document}